**Preseismic oscillating electric field "strange attractor" like precursor, of T=14 days, triggered by M1 tidal wave. Application on large (Ms > 6.0R) EQs in Greece (March 18[th], 2006 - November 17[th], 2008).**

Thanassoulas[1], C., Klentos[2], V., Verveniotis, G.[3], Zymaris, N.[4]

1. Retired from the Institute for Geology and Mineral Exploration (IGME), Geophysical Department, Athens, Greece.
   e-mail: thandin@otenet.gr - URL: www.earthquakeprediction.gr

2. Athens Water Supply & Sewerage Company (EYDAP),
   e-mail: klenvas@mycosmos.gr - URL: www.earthquakeprediction.gr

3. Ass. Director, Physics Teacher at 2[nd] Senior High School of Pyrgos, Greece.
   e-mail: gver36@otenet.gr - URL: www.earthquakeprediction.gr

4. Retired, Electronic Engineer.
   e-mail: nik.zym@tellas.gr

## Abstract.

The "strange attractor like" precursor, calculated from the Earth's oscillating electric field registered at **PYR** and **HIO** monitoring sites located in Greece, is studied in the domain of **T = 14 days**. It is assumed that the generating precursory signals focal mechanism is triggered by the corresponding **M1** (moon declination) tidal wave. The obtained results from the analysis of eight (**8**) cases of large (**Ms>6.0R**) EQs that occurred from March 18[th], 2006 to November 17[th], 2008 suggest the validity of the method. Moreover, it is found that the specific methodology applied for **T = 14 days** behaves very closely to the same one when applied for **T = 1 day** even though there is a resolution decrease in the calculated predictive time window for the occurrence of the oncoming large EQ.

It is speculated that this type of precursor, once it is present in one distinct oscillating component of the seismic precursory generated electric field, then, most probably, it is present in most of its other oscillating components. The latter suggests the investigation of the preseismic precursory electric field at its longer wavelengths i.e. components triggered by the **S_sa (6 months, moon declination)** oscillating components.

The large value of the obtained success rate (predicted EQs / total no. of large EQs) suggests its use as a time prediction tool in the domain of the "short-term prediction".

## 1. Introduction.

After a strong earthquake has occurred at a seismogenic area, the immediate question which arises, is: when will it strike again? The seismologists worked out firstly this question because of its societal significance. This question implies, although in an indirect way, that the seismogenic area is already known and the expected magnitude of the future earthquake is almost similar, in magnitude, to the previous one or strong enough, so that it must be taken seriously into account.

At the early steps of the scientific research for a successful earthquake prediction, the recurrence times of strong earthquakes, at a specific seismogenic area, were analyzed with statistical methods. The obtained results mainly refer to the long-term prediction. The fact that strong earthquakes are not very frequent prohibits increased time resolution of the statistical methods which are used, mainly due to the fact that the "sampling interval" is too long. A different statistical scheme, analyzes the **frequency – magnitude dependence and their logarithmic proportionality, the "b-value"** (Ma 1978, Smith 1986, Molchan et al. 1990) and variations of coda Q (Jin and Aki 1986, Sato 1986). Major earthquakes have been preceded by **seismic quiescence**. Kanamori (1981), Lay et al. (1982), Wyss et al. (1988) reviewed the methodology. Scholz (1988) studied the mechanisms, which could be responsible for this phenomenon, while Schreider (1990) proposed a statistical basis for making reliable predictions, based on quiescence. Gupta (2001) used the same methodology for the medium-term forecast of the 1988 northeast India earthquake. As a result of this inability of the statistical methodologies to provide adequate time resolution (predictive time window for the future strong earthquake), which could be of some use for the society, different methodologies, more sophisticated, were developed. **The algorithm CN** is one of them (Keilis-Borok et al. 1990). This algorithm allows diagnosis of the times of increased probability of strong earthquakes (TIPs). The **CN** stands for the application of the TIPs methodology for California – Nevada, while a version addressed to magnitudes larger than M=8R is assigned the name **"algorithm M8"** (Keilis-Borok and Kossobokov 1990; Romachkova et al. 1998). Varnes (1989) studied the **accelerating release of seismic energy** or seismic moment either to time elapsed, or to time remaining, in the period preceding a main shock. He introduced empirical relations, having origins in both experiment and physical theory, going back many decades. The acceleration process was analyzed in laboratory experiments and was applied before strong earthquakes in Kamchatka and Italy by Di Giovambattista and Tyupkin (2001).

Narkunskaya and Shnirman (1990) proposed the **multi-scale model of defect development**. Following this methodology and through simulation of the lithosphere, to a hierarchical discrete structure, which consists of some singularities, allows one to predict the time of appearance of "large" defects.

According to the **Load-Unload Response Ratio (LURR) methodology**, when a system is stable, its response to loading corresponds to its response to unloading, whereas when the system approaches an unstable state, the response to loading and unloading becomes quite different (Yin et al. 1995, 1996, 2002).

Seismic quiescence and accelerating seismic energy release can be further detailed by the use of **the RTL algorithm**. This algorithm (Sobolev 2001, Sobolev et al. 2002, Di Giovambatista et al. 2004) analyze the RTL (Region, Time, Length) prognostic parameter, which is designed in such a way, to have a negative value if, in comparison with long-term background, there is a deficiency of events in the time – space vicinity of the tested area. The RTL parameter increases if activation of seismicity, takes place.

**Seismicity "Pattern recognition algorithms"** such as "ROC" – range of correlation – and "Accord" were used by Keilis-Borok et al. (2002) to identify premonitory patterns of seismicity, months before strong earthquakes in Southern California. Another version of the pattern recognition methodology takes into account the earthquake intensities, in order to forecast the time of occurrence (Holliday et al. 2006).

In **the RTP (reverse tracing of precursors)** methodology, the precursors are considered in reverse order of their appearance (Keilis-Borok et al. 2004).

**The "Space-time ETAS" methodology** (Ogata et al. 2006) is a further extension and improvement of the seismic quiescence methodology. **ETAS** stand for Epidemic Type Aftershock Sequence.

Yamashina (2006) used the **statistical test of time shift** for prediction studies in Japan.

**The algorithm SSE** (subsequent strong earthquake) was designed for prediction of relatively strong earthquakes following a strong earthquake. A subsequent, strong earthquake can be an aftershock or a main shock of larger magnitude.

The **SSE** algorithm resulted from the analysis of 21 case histories in California and Nevada. Then, it was, retrospectively, tested in 8 seismic regions of the world (Gvishiani et al. 1980, Levshina et al. 1992, Vorobieva et al. 1993, Vorobieva et al. 1994, Vorobieva 1994, Vorobieva 1999).

The statistical treatment of the earthquakes, which occurred at a specific seismogenic area in the past, has been proved more or less sufficient for the "medium-term or long-term" earthquake prediction.

However, in terms of the "short-term" prediction the seismological literature has not even a successful case to present at all. Apart from the long time sampling interval, which elapses between two consecutive strong earthquakes in the same seismogenic area, there is an intrinsic difficulty in using statistical methods towards a successful, short-term earthquake prediction. The latter was analyzed in detail by Thanassoulas (2007).

An entirely different approach towards the short-term earthquake prediction was followed by other researchers. Varotsos et al. (1981) used the **SES** (seismic electric signals), generated short before (in term of days) the occurrence of an earthquake in order to forecast the time of its generation. Details of this methodology were presented by Varotsos (2005). A further development of this methodology is the study of the behaviour of seismicity in the area candidate to suffer a main shock. This area is investigated after the observation of the Seismic Electric Signal activity until the impending main shock (Sarlis et al. 2008; Varotsos et al. 2008). In this method, although its initiation is made by the generation of the SES, the conclusions about the occurrence time of the large pending EQ are the result of the analysis of the, after the SES generation, following seismicity.

A quite different method for the analysis of the Earth's electric field towards short-term earthquake prediction was followed by Thanassoulas (1982), Thanassoulas and Tselentis (1986, 1993). In these studies it was shown that the Earth's electric field starts to oscillate at a period of **T = 24 hours**, triggered by the corresponding tidal components **(K1, P1)** of the lithospheric oscillation, short before the occurrence of the pending EQ. The lithospheric tidal oscillations in return trigger the generation of the oscillating electric field through the activation of large scale piezoelectric mechanisms. A further development of the method was achieved (Thanassoulas et al. 2008a,b), by considering the earth's oscillating field registered by two distant monitoring sites and by applying mapping techniques (Nusse and Yorke, 1998; Korsch, Jodl, and Hartmann, 2008).

The triggering of the large scale piezoelectric mechanism in the lithosphere (Thanassoulas, 2007) has been adopted as the main mechanism of generating preseismic electric signals of any type observed so far (see the particular case of **SES**, Thanassoulas, 2008). Therefore, it is worth to investigate the presence of the **"strange attractor like"** behaviour of the Earth's electric field at its longer wavelength content before the occurrence of any large EQ. In particular, the Earth's electric field will be considered at periods of **T = 14 days**. This specific period of the oscillating electric field corresponds to the **M1** tidal wave, and therefore to the lithospheric oscillation too, caused by the Moon declination.

The aim of this work is:
  a) to investigate the presence of the "strange attractor like" precursor in the recorded Earth's electric field data (of period **T=14 days**), in a rather long time window, in relation to the occurrence time of large EQs **(M >6.0R)** which did occur in the same time window and
  b) to evaluate the predictive capability of the **"strange attractor like"** precursor calculated by the Earth's electric oscillating field at a period of **T = 14 days**.

## 2. The data.

The study period spans for 30 months from March 18[th], 2006 to November 17[th], 2008. This period of time was selected due to the fact that **HIO** monitoring site was established on March 18[th], 2006. Therefore, a pair of monitoring sites of the Earth's electric field was formed with **PYR** monitoring site, having almost similar technical specifications of receiving dipoles (Thanassoulas, 2007) and consequently the **"strange attractor like"** methodology can be applied on the pair of the recorded data. The location of the **PYR** and **HIO** monitoring sites is presented in figure **(1)**.

During this time period eight **(8)** large earthquakes occurred with magnitude **Ms>6.0R**. Their date, depth of occurrence, and magnitude are tabulated in **TABLE – 1.** These EQs are numbered from **(1)** to **(8).** This number denotes the corresponding EQ in the map of figure **(1)**.

**TABLE – 1.**

| Earthquake | | Depth (in Km) | Magnitude (Ms) |
|---|---|---|---|
| 1. March 25[th], | 2007 | 15 | 6.0 |
| 2. January 6[th], | 2008 | 86 | 6.6 |
| 3. February 14[th], | 2008 | 41 | 6.7 |
| 4. February 20[th], | 2008 | 25 | 6.5 |
| 5. June 8[th], | 2008 | 25 | 7.0 |
| 6. June 21[st], | 2008 | 12 | 6.0 |
| 7. July 15[th], | 2008 | 56 | 6.7 |
| 8. October 14[th], | 2008 | 24 | 6.1 |

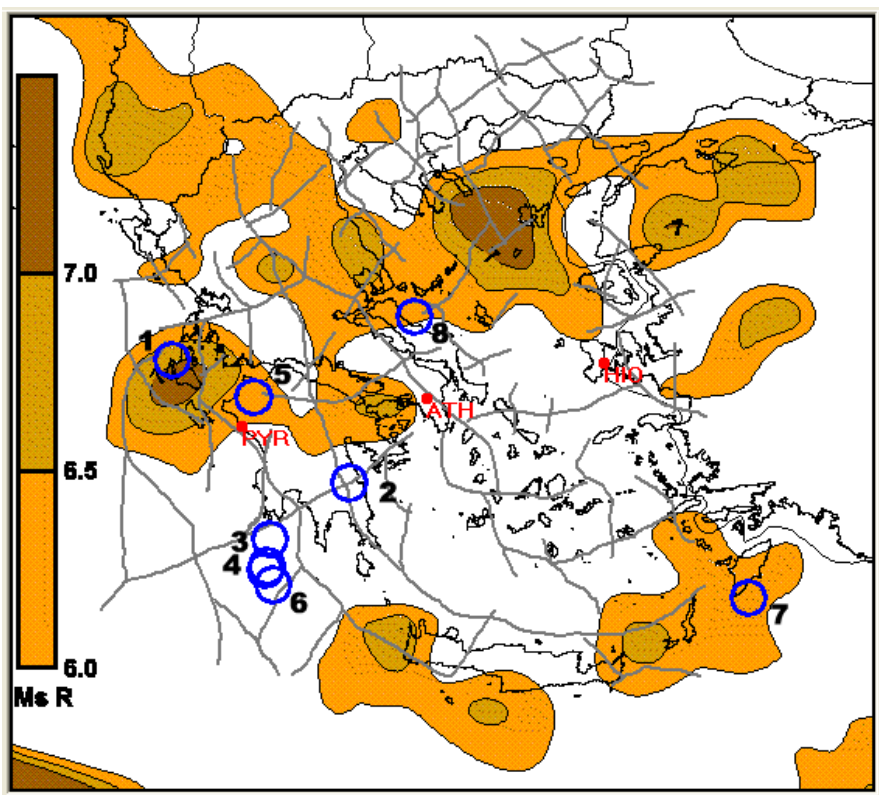

Fig. 1. Location (blue circles) of the EQs of **Ms>6.0R** which occurred in Greece during March 18[th], 2006 to November 17[th], 2008. The numbers refer to **TABLE - 1**. The colour coded areas correspond to areas of large seismic potential while the thick grey lines depict the deep lithospheric fracture zones (Thanassoulas, 2007).

The Earth's electric field which was recorded, during the study period, by the **PYR** monitoring site is presented in figure **(2).** The large EQs that occurred in the same period are denoted by red bars.

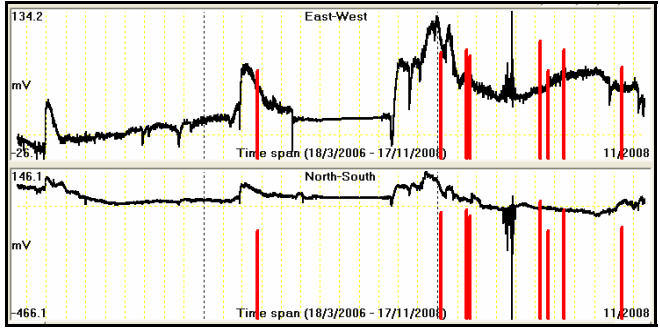

Fig. 2. Raw data (black line) of the Earth's electric field recorded by **PYR** monitoring site during the period from March 18[th], 2006 to November 17[th], 2008. The red bars indicate the occurrence of large (**Ms>6.0R**) EQs.

The data of figure (**1**) were band-pass filtered (**FFT**) with a center filter period of **Tc = 14** days, a bandwidth of 28 days and are presented in figure (**2a**).

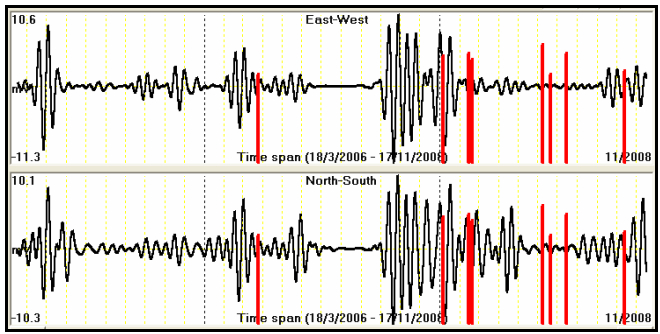

Fig. 2a. Oscillating data (**Tc = 14 days**, black line) of the Earth's electric field obtained from **PYR** raw data (after band-pass filtering) during the period from March 18[th], 2006 to November 17[th], 2008. The red bars indicate the occurrence of large (**Ms>6.0R**) EQs.

The Earth's electric field which was recorded, during the study period, by the **HIO** monitoring site is presented in figure (**3**). The large EQs that occurred in the same period are denoted by red bars.

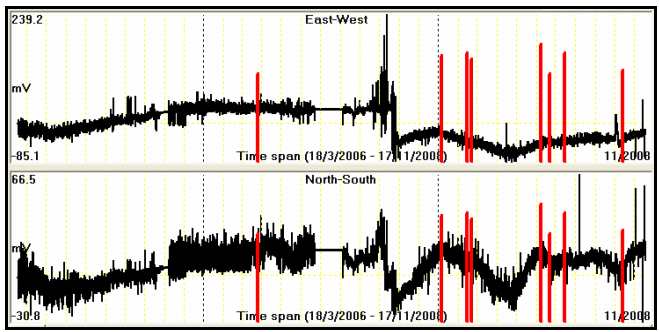

Fig. 3. Raw data (black line) of the Earth's electric field recorded by **HIO** monitoring site during the period from March 18[th], 2006 to November 17[th], 2008. The red bars indicate the occurrence of large (**Ms>6.0R**) EQs.

The data of figure (**3**) were band-pass filtered (**FFT**) with a center filter period of **Tc = 14** days, a bandwidth of 28 days and are presented in figure (**3a**).

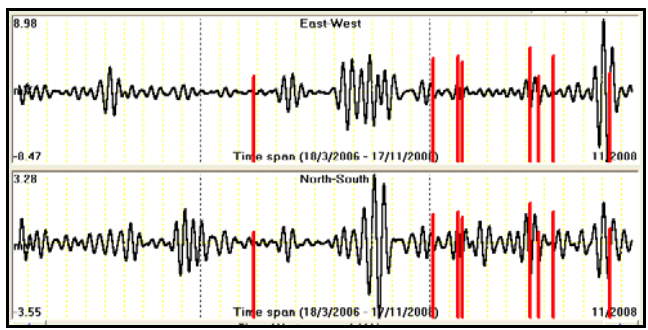

Fig. 3a. Oscillating data (**Tc = 14 days**, black line) of the Earth's electric field obtained from **HIO** raw data (after band-pass filtering) during the period from March 18[th], 2006 to November 17[th], 2008. The red bars indicate the occurrence of large (**Ms>6.0R**) EQs.

The depicted data in figures (**2a**) and (**3a**) were used to compile the corresponding phase maps (Thanassoulas et al. 2008a) for time intervals related to the occurred EQs. The days which were characterized by the presence of ellipses were given a value of (**1**), while the presences of hyperbolas (absence of ellipses) were given a value of (**0**). Therefore, the entire study time period (in days) is characterized by "**0**"s and "**1**"s. It is clear that all along the presence of an ellipses it will characterize the corresponding time period (in days) with "**1**"s. The aim of this study is to compare the time of occurrence of each large EQ to the presence of any ellipse (consecutive "**1**"s) short before, during and after its occurrence. A graph which will present EQ time of occurrence and magnitude against the presence of an ellipse has to present EQ magnitudes and "**1**"s in a normalized form so that comparison is easily facilitated. To this end the time series of "**1**"s and "**0**"s were convolved with the weights of a filter of seven (**7**) days long and with the following weights: **1, 2, 3, 4, 3, 2, 1**. When this filter is applied on a series of consecutive "**1**"s it produces a centre value of (**16**). This value is in excess than the larger magnitude observed of **Ms = 7.0R**. So the final value of the filter weights were converted to (**1, 2, 3, 4, 3, 2, 1**) **/3**. With this modification the output of the filter does not exceed the value of **5.333** and consequently the corresponding graph of EQs time of occurrence – magnitude against the filtered data of presence ("**1**"s) or absence ("**0**"s) of ellipses will be nicely presented. This operation is made clearer from the following examples of real EQs.

## 3. Examples presentation.

For each EQ of **TABLE – 1** the corresponding phase map has been compiled for three (**3**) distinct and consecutive periods of time. The first one corresponds to the period before the occurrence of the EQ while no any "strange attractor like" precursor was present, the second period corresponds to the one short before the EQ occurrence while the "strange attractor like" precursor was present, and the last one corresponds to the period after the occurrence of the EQ while the "strange attractor like" precursor has vanished. In cases when no "strange attractor like" precursor exists at all then only one phase map is compiled for the entire before – during and after the EQ occurrence period of time. The presence of the "strange attractor like" precursor is also presented as a graph vs. magnitude and time of occurrence of the corresponding EQ. Following are presented the results of the analysis of the EQs of **TABLE – 1.**

### 3.1. March 25th, 2007, Ms = 6.0R, Z=15Km

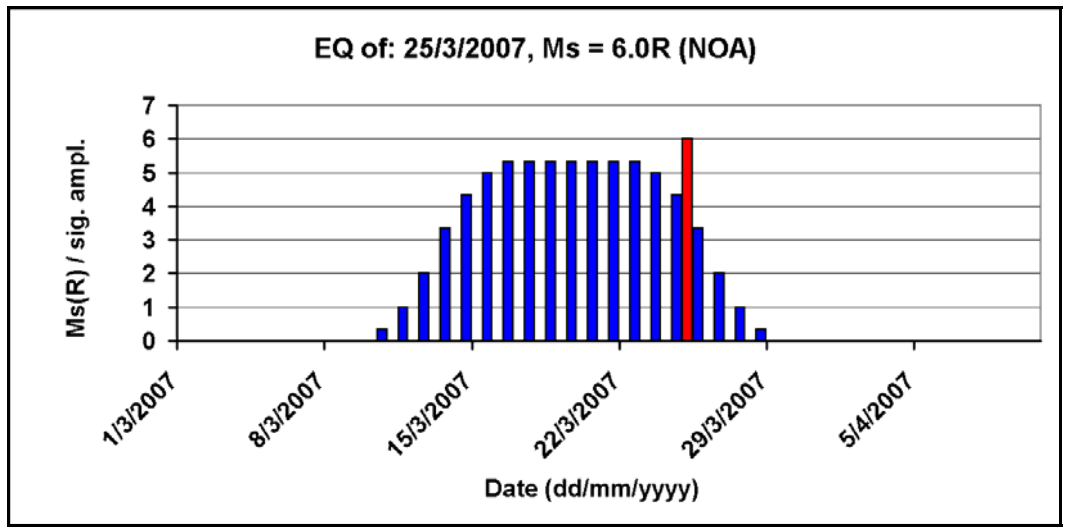

Fig. 4. "Strange attractor like" precursor presence (blue bars) vs. EQ magnitude and time of occurrence (red bar).

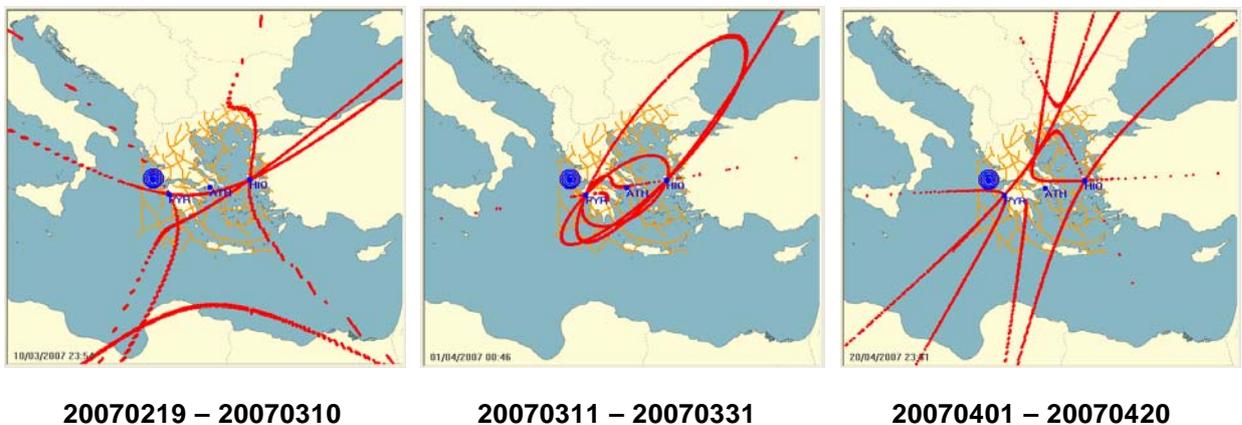

| **20070219 – 20070310** | **20070311 – 20070331** | **20070401 – 20070420** |

Fig. 5. Compiled phase maps for the corresponding periods of time (in yyyymmdd mode). The blue concentric circles denote the EQ location.

### 3.2. January 6th, 2008, Ms = 6.6R, Z=86Km

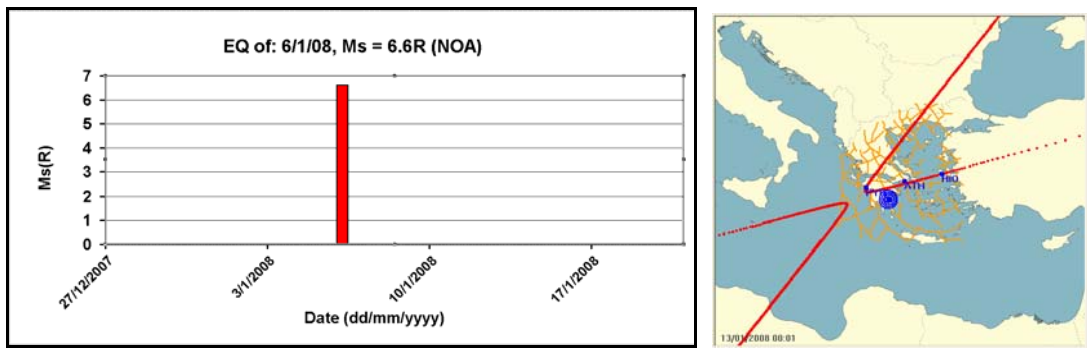

20080101 - 20080112

Fig. 6. "Strange attractor like" precursor absence, EQ time of occurrence and magnitude (left, red bar) and the compiled phase map (right) for the corresponding period of time (in yyyymmdd mode). The blue concentric circles denote the EQ location.

### 3.3. February14th, 2008, Ms = 6.7R, Z=41Km and February 20th, 2008, Ms = 6.5R, Z=25Km

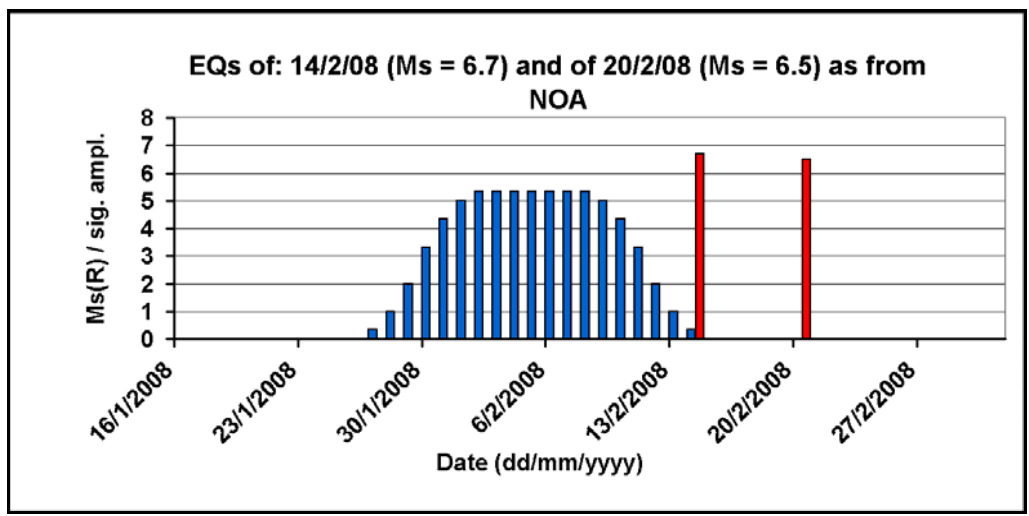

Fig. 7. "Strange attractor like" precursor presence (blue bars) vs. EQs magnitude and time of occurrences (red bars).

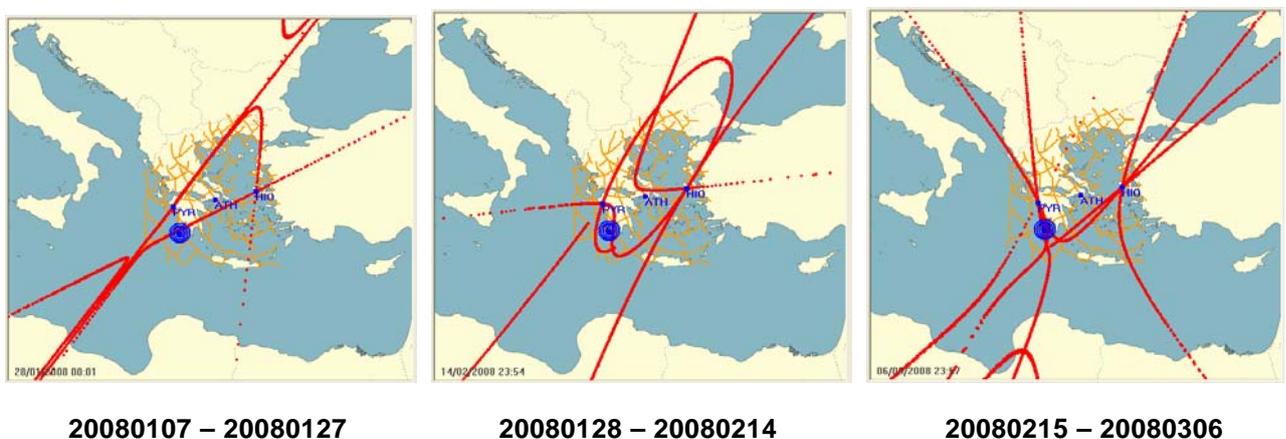

20080107 – 20080127          20080128 – 20080214          20080215 – 20080306

Fig. 8. Compiled phase maps for the corresponding periods of time (in yyyymmdd mode). The blue concentric circles denote the EQs location.

**3.4.** June 8[th], 2008, Ms = 7.0R, Z=25Km and June 21[st], 2008, Ms = 6.0R, Z=12Km

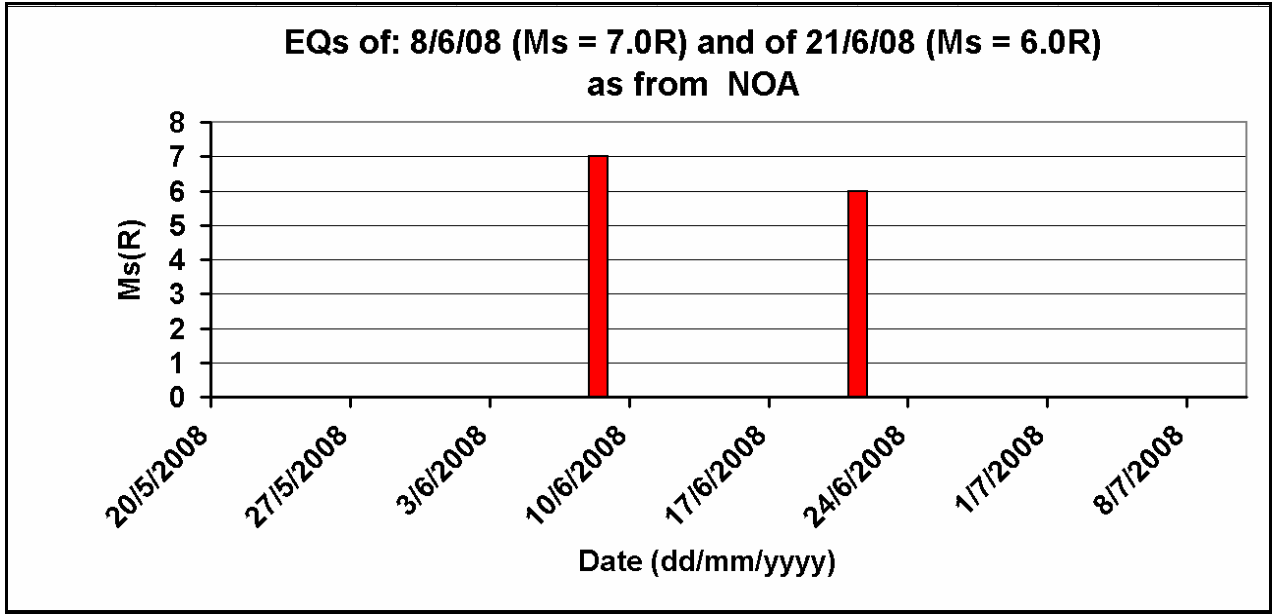

Fig. 9. "Strange attractor like" precursor absence, EQs time of occurrence and magnitude (red bars).

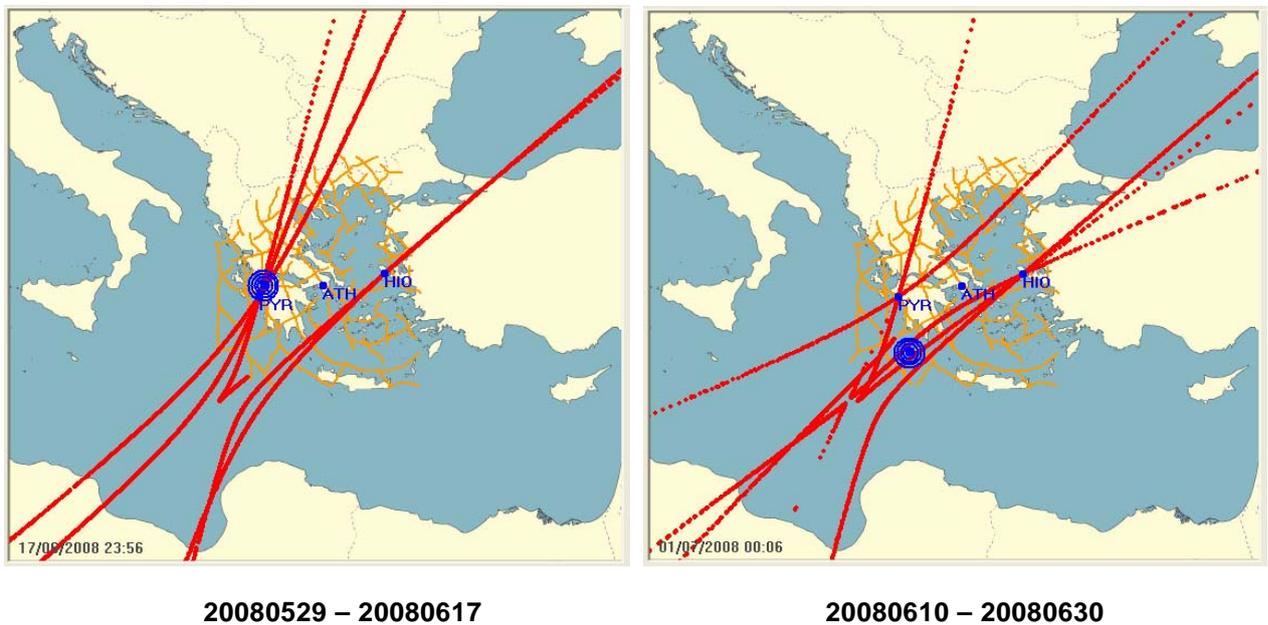

**20080529 – 20080617**          **20080610 – 20080630**

Fig. 10. Compiled phase maps for the corresponding periods of time (in yyyymmdd mode). The blue concentric circles denote the EQs location.

**3.5. July 15th, 2008, Ms = 6.7R, Z=56Km**

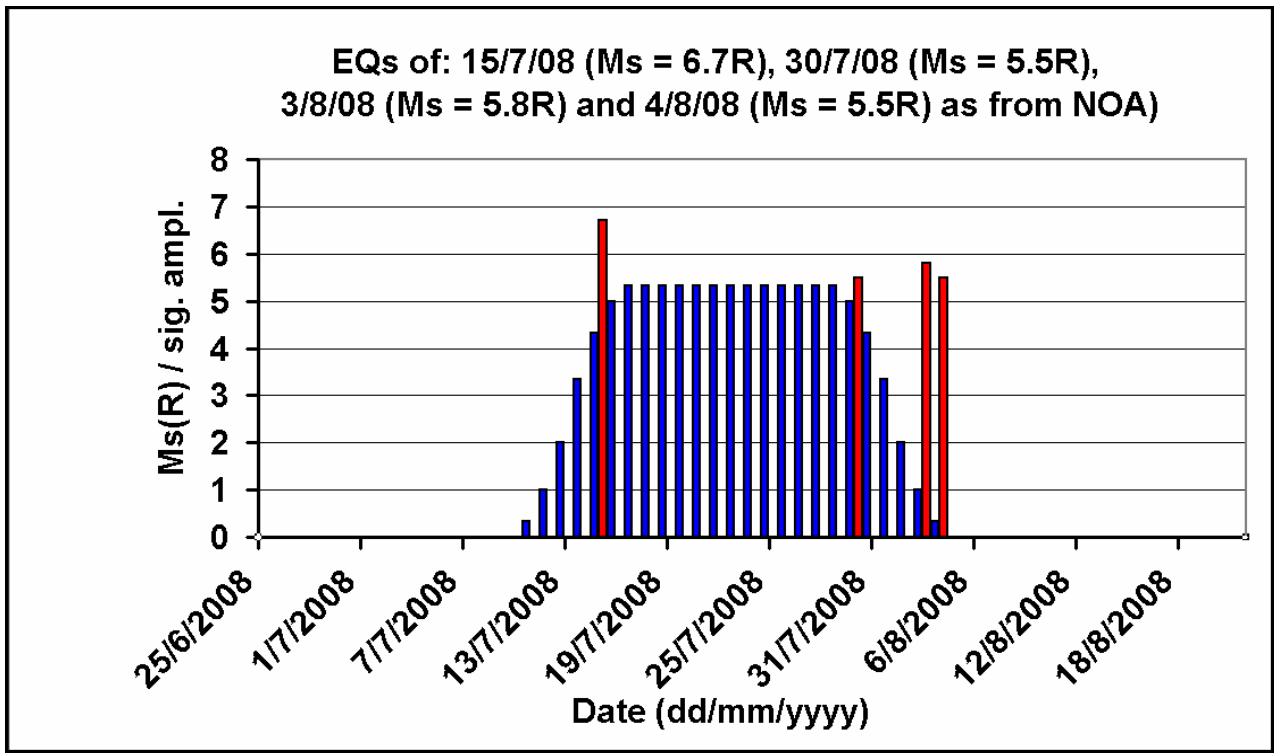

Fig. 11. "Strange attractor like" precursor presence (blue bars) vs. EQs magnitude and time of occurrences (red bars).

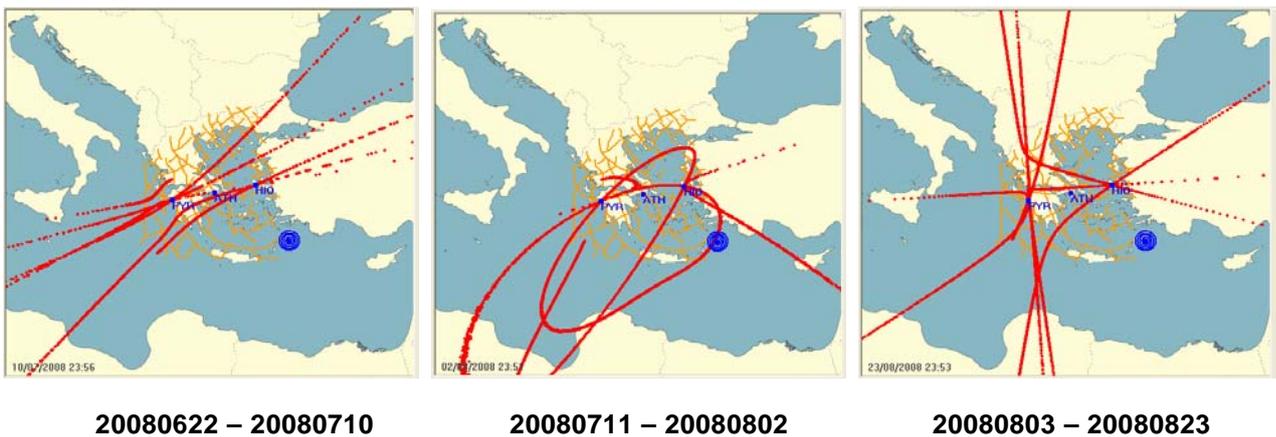

| 20080622 – 20080710 | 20080711 – 20080802 | 20080803 – 20080823 |

Fig. 12. Compiled phase maps for the corresponding periods of time (in yyyymmdd mode). The blue concentric circles denote the location of the EQ of **Ms=6.7R** magnitude.

### 3.6. October 14th, 2008, Ms = 6.1R, Z=24Km

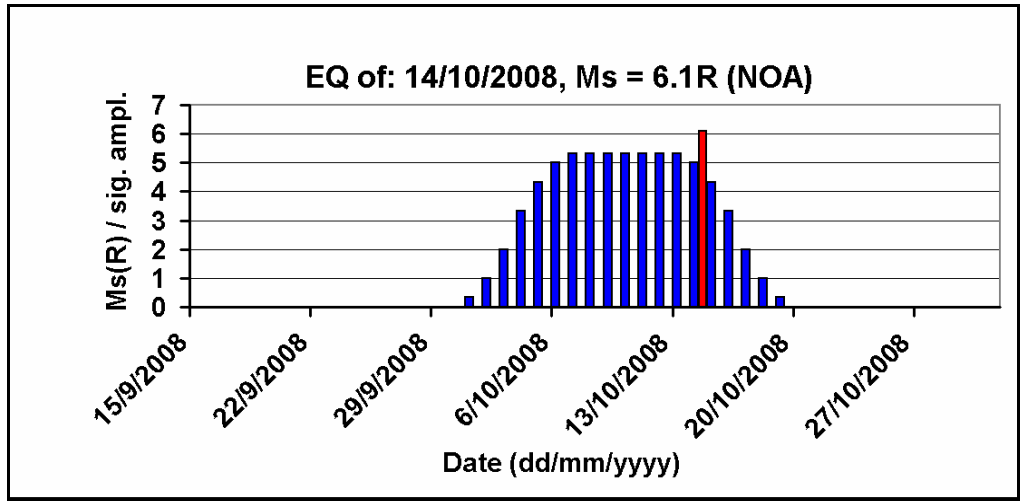

Fig. 13. "Strange attractor like" precursor presence (blue bars) vs. EQ magnitude and time of occurrence (red bar).

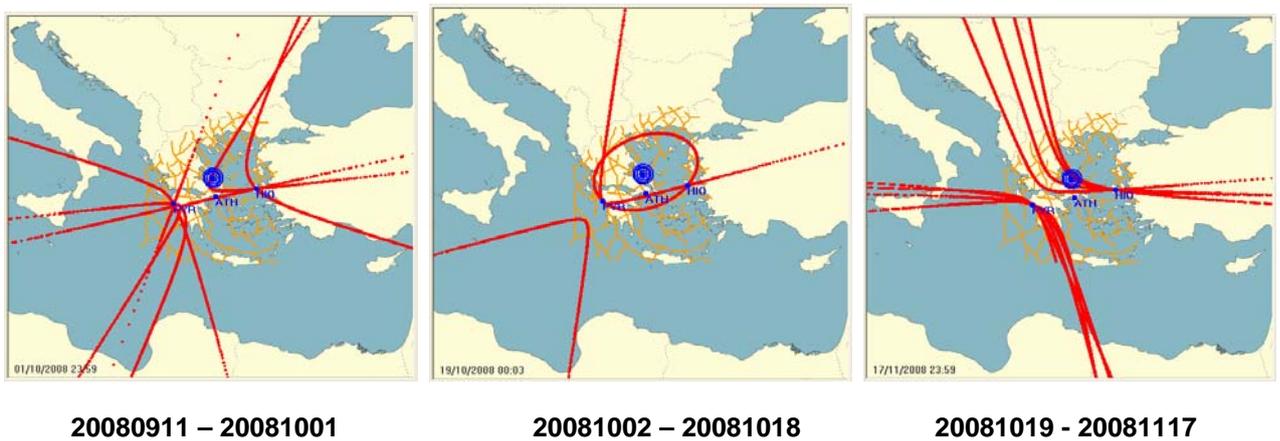

**20080911 – 20081001**    **20081002 – 20081018**    **20081019 - 20081117**

Fig. 14. Compiled phase maps for the corresponding periods of time (in yyyymmdd mode). The blue concentric circles denote the EQ location.

## 4. Discussions – conclusions.

The results obtained from the application of the "strange attractor like" method on the data registered by the **PYR** and **HIO** monitoring sites are tabulated in **TABLE – 2**. Its first column contains the numbering, date, magnitude and depth of occurrence of each EQ. The second column indicates whether a precursor was present (**Y**) or not (**N**). The third column indicates the days when the precursor was initiated before the EQ occurrence time (in days) and the forth column presents the duration of the precursor in days.

**TABLE – 2.**

| Earthquake (date, magnitude, depth) | Precursor presence | Precursor initiation before EQ (in days) | Precursor duration (in days) |
|---|---|---|---|
| 1. March 25th, 2007,     Ms = 6.0R, Z=15Km | Y | 15 | 19 |
| 2. January 6th, 2008,     Ms = 6.6R, Z=86Km | N | - | - |
| 3. February14th, 2008,    Ms = 6.7R, Z=41Km | Y | 19 | 19 |
| 4. February 20th, 2008, Ms = 6.5R, Z=25Km | Y | 26 | 19 |
| 5. June 8th, 2008,        Ms = 7.0R, Z=25Km | N | - | - |
| 6. June 21st , 2008,      Ms = 6.0R, Z=12Km | N | - | - |
| 7. July 15th, 2008,       Ms = 6.7R, Z=56Km | Y | 5 | 25 |
| 8. October 14th , 2008,     Ms = 6.1R, Z=24Km | Y | 14 | 19 |

The average value of the precursor initiation in days is: Precursor initiation = 16 (15.8) days, while the corresponding average value for the precursor duration is: Precursor duration = 20 (20.2) days. Both values are larger than the corresponding ones (precursor initiation = 9 days, precursor duration = 7 days) for **T = 1 day** (Thanassoulas et al. 2008b). The observed difference in both values is attributed to the longer wavelength used in this particular case.

An exceptional case of the studied precursor is the one of figure (**11**). Although the precursor was initiated very short before the corresponding large EQ (**July 15$^{th}$, 2008, Ms = 6.7R, Z=56Km**) its duration was prolonged by the smaller EQs that did occur (**30/7 – 3,4/8/2008, Ms = 5.5R, 5.8R, 5.5R**) just at the end of the precursor.

By taking into account the results presented in the second column of the latter table it is possible to initially evaluate the behaviour of the method in terms of true prediction probabilities. Thus, having eight (**8**) data samples (**D$_{total}$**) as input and a total of five (**5**) predicted (**P$_{predicted}$**) cases, the success rate (**S$_{rate}$**) is defined as:

$$S_{rate} = P_{predicted} \ / \ D_{total} \tag{1}$$

by taking into account the values of: $\quad$ $$P_{predicted} = 5 \quad \text{and} \quad D_{total} = 8 \tag{2}$$

then **S$_{rate}$** takes the value of**:**

$$S_{rate} = 5 \ / \ 8 = 0.625 \ \text{ or } 62.5\% \tag{3}$$

The calculated **S$_{rate}$** value is worth to consider only if it is much larger than the results **S$_{random}$** obtained from a random method used for EQ occurrence time selection. Therefore, **S$_{rate}$** will be compared to the **S$_{random}$** that results from a completely random guess selection for the times of occurrence of five (**5**) EQs for the same period of time. This analysis is performed as follows:

The entire study period of time (March 18$^{th}$, 2006 to November 17$^{th}$, 2008) is divided into 60 consecutive sub-periods of **15** days each. The selection of **15** days length for each sub-period of time was based on two reasons. The first is that it is very close to one (**1**) wavelength (**T = 14 days**) of the **M1** tidal component. The second is that it eases further calculations. Moreover, it is initially assumed that only one (**1**) EQ will occur in the study period. It is clear that this EQ can occur in any one of the 60 considered sub-periods. Consequently, the probability (**P$_1$**) to guess by chance successfully the sub-period of its occurrence is:

$$P_1 = 1/60 = 1.66\% \tag{4}$$

If now it is required to guess the sub-period of occurrence of one (**1**) EQs out of eight (**8**) which occur in the entire study period, then the probability **P$_{1/8}$** value to guess successfully by chance the sub-period of its occurrence becomes eight (**8**) times larger than **P$_1$**:

$$P_{1/8} = 8/60 = 8P_1 \tag{5}$$

Next, let us consider the case when two EQs must be guessed. The first EQ has a **P$_1$** value of **8/60**. The second EQ has a probability value **P$_2$** as:

$$P_2 \ = 7/59 \tag{6}$$

due to the fact that only seven (**7**) EQs and only **59** sub-periods remain to be guessed after the first guess. Therefore, the probability **P$_{2/8}$** to guess successfully by chance two (**2**) EQs is:

$$P_{2/8} = \ (8/60) * (7/59) \tag{7}$$

Following this scheme it is possible to calculate the probability **P$_{5/8}$** to guess successfully by chance five (**5**) EQs out of eight (**8**) in the time span of **60** sub-periods of time of the study period as follows:

$$P_{5/8} = \ (8/60) * (7/59) * (6/58) * (5/57) * (4/56) \tag{8}$$

$$P_{5/8} = \ 0.000010253 \quad \text{or} \quad P_{5/8} = 0. \ 0010253\% \tag{9}$$

The calculated "by chance" probability value of **P$_{5/8}$** of equation (**9**) will be compared to the actual success rate **S$_{rate}$** of equation (**3**):

$$(S_{rate} \ / \ P_{5/8} \ ) = 0.625 \ / \ 0.000010253 = 60957.768 \tag{10}$$

Equation (**10**) suggests that the results obtained from the application of the specific methodology are more than worth to be considered and by no means are "just a coincidence".

From the eight (**8**) studied EQs three of them (**2, 5, 6**) from **TABLE – 2** did not produce any such precursor at all. For two of them (**2, 6**) a rational explanation can be given by taking into account the **p-velocity** (Mueller and Landisman, 1966) and electrical properties distribution (Blohm et al. 1977, Thanassoulas 2007)) in the lithosphere. The no. (**2**) EQ did occur at a depth of 86Km where the crystalline properties of the lithosphere are small while plasticity effects are increased. Therefore, no electric signals are expected to be generated. That had been pointed out by Thanassoulas (2007) who suggested that not all large EQs can be predicted by electric methods, but only the ones generated in the crystalline part of lithosphere that is capable of generating electric fields due to its mechanical deformation. The EQ of no. (**6**) occurred at a depth of 12 Km that is very close to the **Fortsch** boundary where there is a drastic change of the velocity of **P** seismic waves. EQs which occur at a depth of a few Km from the ground surface are not expected to generate strong electric signals. Finally, for the EQ of no. (**5**), since it occurred at a depth of 25 Km, there is not any rational explanation and its behaviour remains a big question mark.

Taking into account the previous paragraph it is possible to recalculate the **S$_{rate}$** with different values for **P$_{predicted}$** and **D$_{total}$** by ignoring EQs of no. (**2**) and (**6**) as follows:

$$P_{predicted} = 5 \quad \text{and} \quad D_{total} = 6 \tag{11}$$

and

$$S_{rate} = 5 / 6 = 0.833 \quad \text{or } 83.3\% \tag{12}$$

Moreover, having shown the presence of the "strange attractor like" precursor before large EQs in the frequency band of **T = 14 days**, it is worth to compare it with the presence, for the same period of time, of the corresponding attractors for **T = 1 day** (Thanassoulas et al. 2008b). The results of this analysis are presented in the following **TABLE – 3.**

**TABLE – 3.**

| Earthquake (date, magnitude, depth) | Precursor presence of T=14 days | Precursor presence of T=1day |
|---|---|---|
| 1. March 25$^{th}$, 2007,      Ms = 6.0R, Z=15Km | Y | Y |
| 2. January 6$^{th}$, 2008,      Ms = 6.6R, Z=86Km | N | N |
| 3. February14$^{th}$, 2008,    Ms = 6.7R, Z=41Km | Y | Y |
| 4. February 20$^{th}$, 2008, Ms = 6.5R, Z=25Km | Y | Y |
| 5. June 8$^{th}$, 2008,        Ms = 7.0R, Z=25Km | N | Y |
| 6. June 21$^{st}$ , 2008,      Ms = 6.0R, Z=12Km | N | Y |
| 7. July 15$^{th}$, 2008,       Ms = 6.7R, Z=56Km | Y | N |
| 8. October 14$^{th}$ , 2008,    Ms = 6.1R, Z=24Km | Y | Y |

An immediate look at **Table - 3** reveals that the **EQ no.2**, which occurred at large (**z=86Km**) depth, did not produce either type of precursor. This fact supports what was proposed earlier concerning the absence of crystalline lithospheric properties at such depths. On the other hand, there is a remarkable similarity in the presence of these types of seismic precursors. The calculated **S$_{rate}$** (ignoring **EQ no.2**) for the case of **T = 1 day** and **T = 14 days** are:

$$T = 1 \text{ day:} \quad S_{rate} = 6/7 = .857 \text{ or } 85.7\% \tag{13}$$

$$T = 14 \text{ days:} \quad S_{rate} = 5/7 = .714 \text{ or } 71.4\% \tag{14}$$

These results can be explained by an electric precursors generating mechanism, in the lithosphere, that is triggered in both periods **(T = 1 / 14 days).** In this case, the well-known tidal waves **(M1, K1, P1)** are the most probable triggering mechanism. Since the triggered electric precursor generating mechanism is unique, its generated electric field components (different oscillating electric fields) will behave in more or less the same way. Therefore, the compatibility of **S$_{rate}$** values calculated at **(13)** and **(14)** is well justified. The latter suggests the future investigation of the generation of "strange attractor like" precursors in the time window suggested by the **S$_{sa}$** (**T = 6 months**, Moon declination) tidal component. If this mechanism is valid then the same "strange attractor like" precursors must exist, in this case too, before large EQs. More generally, this type of precursor's presence effect, probably, holds for the entire electric frequency spectrum generated by an electric field generating mechanism. The triggering mechanism of the tidal waves simply intensifies specific wavelengths of the electric field which correspond to certain mechanical triggering wavelengths.

The use of different wavelengths in the calculation of the "strange attractor like" precursor has as an effect the change of the resolution obtained in the calculation of the time window in which a large EQ will occur. In the case of **T = 1 day** it was shown that there was a distinct initiation and duration in days of the precursor before the EQ occurrence (Thanassoulas et al. 2008b). This was not observed for the **T = 14 days** case, on the contrary, the resolution capability decreased.  On the other hand, in the **T = 14 days** case, shorter wavelengths electric interference, which could accidentally modify the used data, is filtered out. Therefore both wavelengths have their drawbacks and merits.

Finally, it can be said that the application of the "strange attractor like precursor", in the domain of **T = 14 days**, is a method that can provide valuable information concerning the time of occurrence of an oncoming large EQ. Therefore it can contribute a lot towards the "short-term earthquake prediction".

# 5. References.